# Sensory Polymorphism and Behavior: When Machine Vision Meets Monkey Eyes

Satohiro Tajima[1,*]

1) Department of Basic Neuroscience, University of Geneva. CMU, 1 Rue Michel-Servet, 1211 Genève, Switzerland.
*satohiro.tajima@gmail.com

## Abstract

Polymorphism in the peripheral sensory system (e.g., congenital individual differences in photopigment configuration) is important in diverse research fields, ranging from evolutionary biology to engineering, because of its potential relationship to the cognitive and behavioral variability among individuals. However, there is a gap between the current understanding of sensory polymorphism and the behavioral variability that is an outcome of potentially complex cognitive processes in natural environments. Linking peripheral sensor properties to behavior requires computational models of nervous processes transforming sensory representations into actions, based on quantitative data from physiological and behavioral studies. Recently, studies based on machine vision approaches are shedding light on the quantitative relationships between sensory polymorphisms and the resulting behavioral variability. To reach a convergent understanding of the functional impacts of sensory polymorphisms in realistic environments, a close coordination among physiological, behavioral, and computational approaches is required. Aiming at enhancing such integrative researches, this paper provides an overview of the recent progress in those interdisciplinary approaches, and suggests effective strategies for such integrative paradigms.

## Background: what is it like to see with different color vision?

Human and animal visions in many cases have polymorphisms In particular, color perception in humans can be markedly different among individuals partially due to differences in the expression of color-sensitive photopigments, because of genetic polymorphisms (Neitz et al., 1996; Neitz and Neitz, 2011). The majority of humans are trichromatic: i.e., light spectra are encoded by the population activity of long-wavelength (L)-sensitive, middle-wavelength (M)-sensitive, and short-wavelength (S)-sensitive cones. Observers lacking one class of photopigments (dichromatic observers) are insensitive to the difference between a certain pair of colors (e.g., red and green), which are sometimes called "confusing colors.

The polymorphism in early visual systems is an intriguing problem for science and our daily lives. For example, seeing colors is crucial for finding fruits in wild environments as well as for recognizing traffic signals, and the polymorphisms in color vision directly affect the performances in those tasks. Lacking skills in discriminating particular color sets sometimes cause difficulties in daily lives. Although gene therapies under development may enable us to manipulate color vision in the near future (Mancuso et al., 2009), there are currently few clinical methods to "cure" the dichromatic observers to have sensitivity to the difference between confusing colors. An alternative strategy to compensate for dichromatic vision from an engineering viewpoint is to design visual materials with sufficient information to be perceived by dichromatic observers (Ichikawa et al., 2003; Rasche et al., 2005; Huang et al., 2008; Machado et al., 2009). In practice, however, the compensatory design is not necessarily straightforward because it requires prediction of the mechanism of perceiving images in dichromatic observers. Such a prediction is particularly difficult for complex visual scenes, where the overall percepts of scenes are organized not only by color but also by multiple different features including luminance contrast, shape, textures and motion. Therefore, estimating the behavioral impacts of polymorphic color vision is not only an issue with cones at retinal level, but also requires understanding of the central nervous system and the of the structure of the external visual environment.

Originating from ophthalmology and visual science, the investigation of polymorphic color vision has expanded to diverse research fields during the last century. Evolutionary ecologists have discussed the advantage of trichromacy in the natural environment. However, the comparison between subjective percepts in individuals having different sensory conditions also involves metaphysical issues (including questions such as "what qualia is experienced by two persons with different color-visions?"), which have interested philosophers for long times (Clark, 1985). Nevertheless, as we will see in the subsequent sections, certain computational techniques that are originated from machine vision contexts allow us to compare the visual percepts among different sensory conditions. This short review article summarizes the recent progresses in researches linking sensory polymorphisms to perception and behavior in complex visual scenes in natural environments. The aim of this article is to enhance the interdisciplinary collaboration among investigators using physiological, behavioral, and computational approaches by associating the progresses in multiple research fields from a consistent viewpoint, and by suggesting a general framework for obtaining a convergent understanding based on behavioral studies guided by recently developed computational techniques. In the subsequent sections, I first review the recent approaches to model the color vision polymorphisms in wiled monkeys and human. In particular, three different approaches are introduced: (i) direct assessment of colorimetric properties combined with machine learning techniques, (ii) computational color substitution, and (iii) visual salience models reflecting sensory polymorphisms. Then, I discuss a general framework for investigating individual differences due to sensory polymorphisms by combining behavioral paradigms and machine vision methodologies.

## Ecology of polymorphic color vision

Many primates, including human, have trichromatic color vision, which enables them to discriminate a much wider range of hues compared to dichromatic mammals including their ancestors. This characteristic visual ability in primates has been raising a long-lasting question concerning the evolution of enhanced color vision in the human lineage (Allen, 1879; Mollon, 1989; Jacobs, 2010). The primate trichromacy, which is sensitive to the spectral difference in long-to-middle (L/M) wavelengths, is widely considered to be a consequence of evolutionary adaptation, in particular for finding reddish ripe fruits in greenish foliage (Mollon et al., 1984; Mollon, 1989; Osorio and Vorobyev, 1996; Tan and Li, 1999; Dominy and Lucas, 2000; Regan et al., 2001; Smith et al., 2003; Sumner and Mollon, 2003; Surridge et al., 2003). At the same time, researchers have been puzzled by the profound inter- and intra-specific variability in primate color vision (Surridge et al., 2003).

Most of the conspecific diversity in color vision is caused by a duplication and divergence of the X-chromosome opsin genes. In particular, the expression of opsin genes encoding L/M opsin is highly variable not only between but also within species. Natural selection models (Osorio and Vorobyev, 1996; Tan and Li, 1999; Dominy and Lucas, 2000) predict overall superiority of trichromats over dichromats. However, surprisingly, trichromatic individuals have not been found to have higher fitness over dichromatic group members in wild populations (Fedigan et al., 2014). Behavioral studies indicate a highly task-dependent nature of the superiority of trichromacy. Although captive studies have revealed the advantages of trichromacy in specific perceptual tasks (Caine and Mundy, 2000; Smith et al., 2003; Pessoa et al., 2005), ethology in the wild environment has even demonstrated cases in which dichromats are superior to trichromats at detecting foods, such as camouflaged insects (Melin et al., 2007, 2010). Therefore, the ecological impacts of each phenotype under natural conditions remain to be clarified.

## Computational impact of polymorphism in natural environments

Recently, Melin et al. (Melin et al., 2014) introduced a machine learning approach to model the visual ability of new world monkeys (*Cebus capucinus*) expressing dichromatic and trichromatic phenotypes to detect foods in the natural environment. By training an artificial classifier (support vector machine) based on cone responses predicted from the spectral reflectance of objects in a wild tropical forest (**Figure 1a**), they estimated how effectively the colors of dietary fruits are discriminated from those of leaves. The classifier modeled with the trichromatic phenotypes was found to correctly discriminate ~75% of the total fruits from background leaves. In contrast, models of dichromatic phenotypes discriminated only <30% of the total dietary fruits. In particular, fruit species having greenish or brownish hues were not detectable by dichromat models. In addition, a significant superiority of trichromacy was found for fruit species with small patch sizes. These results are concordant with the hypothesis that long-distance detection of fruit patches, which provide a high finder's reward, can exert a selective pressure on trichromacy in primates. However, the reason for maintained polymorphic color visions despite the apparent ecological advantages of trichromacy are unclear.

Although the contrast in cone responses constrains the visual information at the retinal level, it is not the only factor that modulates cognition and behavior. The perception of a visual scene is highly dependent on the spatial structure of visual information. Hence, to account for the perceptual relevance of color vision polymorphism in complex scenes, we need to consider not only the local cone contrast but also its relationship to the surrounding context. The next section reviews the recent progress in analyzing the impact of color vision polymorphism under structured visual contents, and explains how a computational model of bottom-up visual attention explains the resulting perceptual divergence.

## Computational substitution of color vision

In early attempts to capture the polymorphic experiences in complex visual scenes, several graphics techniques were proposed to mimic the experiences of dichromats on a trichromatic display (Viénot et al., 1995; Brettel et al., 1997). This was achieved by substituting each set of indistinguishable colors (Pitt, 1944; Thomson and Wright, 1953; Walraven, 1974) with a single representative color chosen to reduce visual discrimination by trichromat observers (**Figure 1b**). The procedure converts the original three-dimensional color space to a two-dimensional space that simulates the color gamut in dichromatic vision. It allows trichromat observers to guess what type of color information would not be accessible for dichromats in complex visual scenes, by comparing the simulated and the original images. This substitution approach resulted in a variety of applications (Ichikawa et al., 2003; Rasche et al., 2005; Huang et al., 2008; Kuhn et al., 2008). A related method is used in behavioral ecology (Melin et al., 2013).

Although the substitution approach is appealing in that it provides precise chromatic metamers for dichromats, it has some weaknesses. In particular, interpreting the simulated colors on the trichromatic display can be problematic because the selection of representative colors is not uniquely determined but contains ambiguity that depends on the algorithm used. For example, red, orange, yellow or green can be used to simulate the appearance of those colors in red–green dichromat vision. For the same reason, the chromatic contrast for trichromats viewing those simulated images contains ambiguity depending on the algorithms. In addition, analysis of the images often relies on subjective impression of the observers in the substitution approach (because the quantitative analysis may suffer from the ambiguity of substitution).

## Perceptual salience as a common metric of the behaviorally relevant visual information

As a solution to the problems in color substitution approach, Tajima and Komine (2015) recently proposed a machine vision approach. They introduced visual salience as a common metric with which perceptual differences due to polymorphic color vision are quantified (**Figure 1c**). The salience is a concept introduced to explain visual attention and developed in the field of computational vision (Treisman and Gelade, 1980; Ullman and Sha'ahua, 1988; Itti and Koch, 2001; Zhaoping, 2002; Bruce and Tsotsos, 2006; Harel et al., 2007; Zhang et al., 2008). Computational models of salience have been reported to predict bottom-up visual attention in trichromatic human observers, including saccadic eye movements (Lee et al., 1999; Itti and Koch, 2001; Bruce and Tsotsos, 2009; Yoshida et al., 2012). However, there have been few studies examining salience models in observers lacking trichromatic color vision. Tajima and Komine (Tajima and Komine, 2015) introduced a hypothetical salience model for non-trichromatic observers, and confirmed a positive correlation between differential salience and the divergence of psychophysical conspicuity judgments among observers with distinct color-vision types. Notably, this approach directly computes salience distributions in an image for different color-vision types, bypassing the simulation of dichromatic vision on a trichromatic display. Moreover, the divergence among color-vision types was ameliorated by manipulating the visual stimuli to reduce the salience difference among observers. It established the causal relationship between modeled salience in various color-vision types and actual perceptual judgments. It is also shown that, in their stimulus set, the inter-individual difference was not significantly correlated to the predictions by other simpler models based on LMS or RGB color contrast, thus the contribution of visual salience (which features multi-scale center-surrounding antagonism and subsequent activity normalization) seems crucial (Tajima and Komine, 2015).

In contrast to the cone-response based characterization (Melin et al., 2014), the salience-based model predicted the case in which dichromats are likely to direct attention to where it is not salient for trichromats, consistent with the natural animal behavior (Caine and Mundy, 2000; Melin et al., 2007, 2010; Smith et al., 2012). The superiority of dichromats in specific tasks is observed also in human psychophysical experiments (Judd, 1943; Newhall et al., 1943; Saito et al., 2006). These findings indicate the potential of salience-based approach to quantitatively bridge the color vision polymorphism and the behavioral divergence in humans and animals, as well as to provide effective applications in visual content design. In particular, a set of visual stimuli that are physically different but matched in terms of visual salience are interpreted as "attentional metamers" (Tajima and Komine, 2015), in an analogy to the metamers in color perceptions (Wyszecki, G. & Stiles, 1982).

The constructing metamers in terms of attentional guidance is a promising framework to discuss the universality of visual information, when no common chromatic metamer exists because of the polymorphism of peripheral sensory organs. However, the salience-based approach does not perfectly explain all the cognitive aspects. One major caveat of the salience-based method is that it does not take into account the effects of higher-level cognition, such as memory-guided object recognition, which is an important extension of the machine vision approach. In particular, recent progresses in machine-vision based image synthesis enables us to generate "metamers" in various perceptual domains other than the color perceptions or attention, textures or natural scenes including complex objects (Portilla and Simoncelli, 2000; Freeman and Simoncelli, 2011; Gatys et al., 2015; Theis and Bethge, 2015). Those advance image synthesis techniques could be used as tools combined with behavioral paradigms to probe how the sensory polymorphisms impact the individual differences in cognitive performances in realistic environments. The subsequent sections discuss how the current machine vision-based techniques could be generalized to a framework for investigating a variety of individual differences in cognitive properties beyond the visual salience.

## A generalized framework for comparing individual perceptual spaces

Returning to the root problem of sensory polymorphisms, the major challenge is that comparing the perceptual contents between two or more observers is generally not straightforward when those observers have different sensory machineries. It is because we have no clear common references for their subjective perceptual spaces when the early sensory representations differ between the observers. For example, two observers having distinct cone properties do not share the visual representations even at the retinal level, thus we cannot assume a common visual metric (e.g., color spaces) for them. Therefore, we need to define a common metric to compare their perceptual spaces from a different viewpoint, which does not rely on the early sensory machineries. The aforementioned visual salience is an example for such a perceptual metric, which is defined in terms of attentional focusing. Here I propose that other functionally defined values in general, which are computed from visual stimuli using machine vision techniques, could be utilized as common metrics to compare the perceptual spaces between observers who do not share the early sensory representations.

**Figure 2** illustrates a general framework for comparing individual perceptual spaces using functionally matched stimuli as references. First, each visual image is mapped to a state in the high-dimensional perceptual spaces of individual observers (**Figure 2a**). We consider an image set ($x$, $y$, and $z$) comprising different versions of images ($x, y, z$). The relationships among those images in an observer's perception are represented by a "constellation" made by those images in the perceptual space. Suppose that there is a sensory polymorphism between the two observers, 1 and 2, such that they have distinct properties in the early sensory system. For example, we can assume that observer 1 has a common trichromatic vision whereas observer 2 has dichromatic vision (e.g., deuteranopia, in which the cones corresponding to the middle wavelength are missing). Due to the difference in the early sensory representations, the constellations in their perceptual spaces are generally different from each other: $(x_1, y_1, z_1) \neq (x_2, y_2, z_2)$, where the subscripts indicate the states in each observer's perceptual space. Note that, in general, even the coordinates for those perceptual spaces are not shared between the observers because the sensory representations are different from the beginning of the visual processing.

To investigate how the polymorphism between the two observers affects their cognitive performances, we can utilize a set of visual stimuli such that some of stimuli are matched between the observers in terms of the perceptual dimensions that we are interested in (the "dimensions of interest" in **Figure 2b, c**). For example, a simplified case of the visual salience is depicted in **Figure 2b**. Let $f$ be the function that computes the salience map of each visual stimulus. We can naturally expect that a metameric image ($y$, a dichromatic image that a particular type of dichromatic observers cannot discriminate from the original trichromatic image) would be similarly perceived also in terms of salience between the observers with different color visions, i.e., $f(y_1) = f(y_2)$, because both the observers compute the luminance and color contrasts based on effectively the same dichromatic information in such an image. Similarly, due to the fact that the visual salience map has lower dimensions than the retinal representations (which usually comprises multiple cone responses), we can generate other "functionally matched stimuli" ($z_1$ and $z_2$) such that they differ from each other in terms of retinal representations yet matched when projected to the dimension of salience ($f(z_1) = f(z_2)$). Note that those metameric images are not necessarily matched in all the perceptual dimensions ("other perceptual dimensions" in the figure); for example, the overall subjective impressions of scene colors themselves could differ between the two observers despite they are matched in terms of the visual salience. Such functionally matched stimuli could be generated, for instance, by reweighting the opponent color signals appropriately (Tajima and Komine, 2012, 2015). We can also test whether the generated stimuli are indeed matched in the dimension of interest by asking whether particular perceptual differences are matched between the two observers (see the figure legend for details). If they do, we can quantify the difference in perceived salience for the stimulus $x$ between the two observers by using those matched stimuli as references.

We can apply the same logic to arbitrary perceptual properties instead of the visual salience (**Figure 3**), such as the quality of material surfaces (Nayar and Oren, 1995; Fleming et al., 2003; Motoyoshi et al., 2007; Nishio et al., 2012), scene categories (Oliva and Schyns, 2000; Oliva and Torralba, 2001), or facial expressions (Dakin and Watt, 2009). What is challenging, however, is to generate the functionally matched stimuli because those high-level perceptual properties generally require the aforementioned function $f$ to take multiple stimulus features including higher-order statistics as their inputs. The complexity of required image features makes the synthesis of functionally matched stimuli harder than in the case of visual salience. Nonetheless, recent advances in computational image synthesis are providing us the way to generate well-controlled images that differ from each other at the pixel-wise representation yet the same or similar in particular aspects of perception. Most of those techniques take "synthesis-by-analysis" approaches by borrowing the idea from machine visions to evaluate the generated images (Portilla and Simoncelli, 2000; Freeman and Simoncelli, 2011; Gatys et al., 2015; Theis and Bethge, 2015). For example, we can synthesize an artificial texture by starting from a noise image and update it iteratively to match the image statistics to those in a reference natural image (Portilla and Simoncelli, 2000). A more recent algorithm allows us to utilize the representation of hierarchical convolutional neural networks as statistics to analyze the images to generate synthetic images to match the neural network responses to complex natural scenes containing multiple visual objects (Gatys et al., 2015) (**Figure 3b**).

Provided those image synthesis methods, we can generate images that are functionally matched between different observers. To do so, we can utilize the outputs of the "analysis" parts in those "synthesis-by-analysis" algorithms. The "analysis" parts, by construction, predict/quantify the perceived properties of given stimuli. By including two or more different models for the "analysis" parts in parallel to reflect the polymorphisms of the early sensory systems, we can predict the individual differences in stimulus perception as well as generate the functionally matched stimuli that share the perceptions among different observers when focused on a specific aspect of cognition (e.g., salience, textures etc.). In addition, whether the perceptions of stimuli are matched between observers or not could be empirically confirmed based on the behavioral paradigm to measure the perceptual differences within each observer, as done in the study of visual salience.

To enable these approaches, we need to integrate a wide variety of methodologies from physiology (to construct the perceptual models under sensory polymorphisms), machine vision (to evaluate and synthesize the stimuli), and behavioral science (to test the model predictions). This would require not only developing recognition and synthesis algorithms for complex tasks, but also close collaborations among researchers from different backgrounds. Recent methodological and conceptual advances are allowing us to step toward achieve this goal. The next section summarizes a general strategy to bridge those different fields to reveal the cognitive/behavioral impacts of sensory polymorphisms.

## Strategies to bridge sensory polymorphism to behavior

The study of color vision and visual salience or foraging behavior in primates provide a good model of a strategy to link sensory polymorphism to resulting behavior. As we have seen in the previous section, this framework can be extended to general visual tasks, such as object recognition or material perception. In addition, this framework would not be limited to color vision: individual variability in the sensory system is observed also in sensory modalities including the variability in visual acuity due to cataracts or macular degeneration and age-related degradation of auditory spectral sensitivity (the notion of salience has been extended to auditory domain (Kayser et al., 2005)). Beyond color vision, the machine vision-guided approaches can be generalized to the study the cognitive and behavioral impact of polymorphism in the sensory system in a variety of sensory domains. The use of machine vision techniques to probe the individual differences in cognitive properties can be classified into three steps as follows:

> *Step 1: modeling.* The first step is to build quantitative models of perception under different sensory conditions. This step requires computational approaches as well as intensive physiological insights about the studied modality, including variability among individuals. In addition, we require the measurement of natural statistics if behavior in natural environment is of particular interest. (For example, to study color-vision polymorphism on scene recognition, researchers can build models to predict the accuracy of object classification based on a concrete algorithm (Lowe, 1999; Fei-Fei and Perona, 2005; Oliva and Torralba, 2007; Bay et al., 2008). Combining those models with the quantitative data in color representation in the early nervous system, they can quantify the influence of color-vision polymorphism on scene recognition.)

> *Step 2: correlation.* The second step is to correlate model predictions and actual behavior. This may requires a new paradigm for behavioral experiments. (For example, when we have different stimuli of visual object, the color vision can impact differently across individual stimuli. Researchers can quantify the plausibility of assumed model by asking whether the model correctly predicts the variability across stimuli.)

> *Step 3: manipulation.* The third step is to manipulate behavior based on model prediction to examine the causal relationship between the hypothesized model and perceptual or behavioral mechanisms. This step may include iterative interactions between behavioral measurement and computational optimization of stimuli. (In the case of scene recognition, researchers can manipulate the visual stimulus of object so as to mitigate or exaggerate the influence of color-vision on the scene classification performance. By examining whether this expected effects of manipulation is observed in human behavior, they can test the causality of modeled mechanisms.)

## Implications and future directions

To reach a convergent understanding about the relationships between sensory polymorphism and behavior, we need to integrate quantitative models of different stages of the cognitive process (**Figure 3a**). This integrative approach is expected to yield insights in a divergent research fields, ranging from basic findings in sensory processing and behavioral ecology, to clinical and engineering applications. The following are examples of potential impacts and remaining issues to be solved.

1) *Systems neuroscience*

    Studying sensory processing under the influence of polymorphism leads to an understanding of the universal mechanism of visual recognition. Visual salience is a critical determinant of bottom-up attention, the first bottleneck through which humans recognize the visual world. However, the precise mechanisms of visual recognition are not well understood the observers with non-common trichromatic color vision. Further quantification is required at the level of photoreceptors (Neitz et al., 1999; Carroll et al., 2004) as well as at high-level cognition such as the categorization of colors (Wachtler et al., 2004).

2) *Machine vision and cognitive sciences*

    Polymorphism provides a clue to understand the functional impact of specific sensory machinery. For example, the role of color in pattern recognition has been of great interest in the context of machine vision (Gevers and Smeulders, 1999; Oliva and Schyns, 2000). Study of polymorphic color visions in humans and animals should hint at the relevance of color information in natural behavior. However, most of the current models of visual polymorphism do not account for the effects of high-order statistics or semantic context on perception. Because the extraction of visual information is guided not only by low-level saliency (bottom-up attentional processes) but also by top-down knowledge of the external world including high-order statistics (Lowe, 1999; Fei-Fei and Perona, 2005; Oliva and Torralba, 2007; Bay et al., 2008), more advanced techniques from machine vision are required to model such high-level visual recognition mechanisms (**Figure 3b**).

3) *Ethology*

    In natural environments, dichromats can be superior to normal trichromatic observers. A series of studies suggests that dichromats are better at finding camouflaged objects against a

multicolored background (Anonymous, 1940; Judd, 1943; Morgan et al., 1992; Saito et al., 2006), a perceptual ability that is critical for foraging, hunting, and avoiding predators in the wild environment (Melin et al., 2007, 2010). Quantitative models that fit such dichromacy as advantageous are yet to be developed. To account for realistic wild behavior, the models should involve viewing distance (Párraga et al., 2002), multisensory integration (Hiramatsu et al., 2009), and social interaction among individuals (di Bitetti and Janson, 2001) .

4) *Clinical study and engineering*

The models linking sensory polymorphism and behavior can be utilized for accessible design of visual content to avoid unintended asymmetry in the information received by observers. The methodology to quantify the perceptual divergence is expected to enhance the effectiveness in creating information graphics (Tajima and Komine, 2012) and prosthetic tools.

## Conclusion

Polymorphism in the sensory system has been studied in diverse research fields. However, bridging sensory polymorphism and behavior requires convergent understanding based on a broad collaboration among researchers adopting physiological, behavioral, and computational approaches. Recent studies in color vision provide good models of such interdisciplinary approaches. The advanced machine vision techniques combined with behavioral experiments can provide tools for those studies. Revealing the relationship between sensory polymorphism and behavioral variability would yield rich insights into the sensory system and behavioral ecology, as well as technological applications to compensate for individual differences in our daily lives. Now is the predawn of such integrative researches.

## Acknowledgement

I thank Chihiro Hamamatsu for comments on the manuscript.

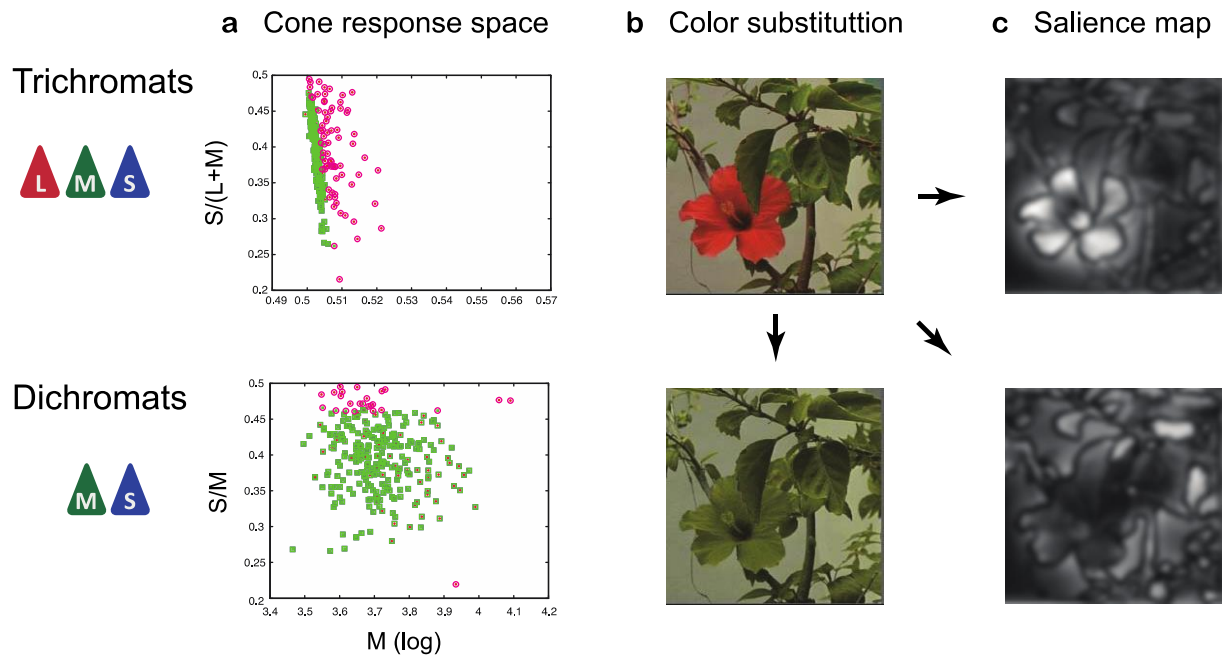

**Figure 1**

Different approaches for quantifying the functional impact of color-vision polymorphism.

(a) Color patch distribution in the cone response spaces. In the trichromat space (top), the fruits (magenta circles) are better dissociated from the leaves (green squares) that in the dichromat color space. Modified form Melin et al. (2012) (Melin et al., 2014).

(b) An example of the cone-model-based color substitution. The visual experience in dichromat (bottom) is mimicked by substituting the confusing colors in the original image (top) with a single color (Viénot et al., 1995; Brettel et al., 1997).

(c) Salience maps based on the cone response models for the trichromat and dichromat observers (Tajima and Komine, 2015). The brightness corresponds to the visual salience value.

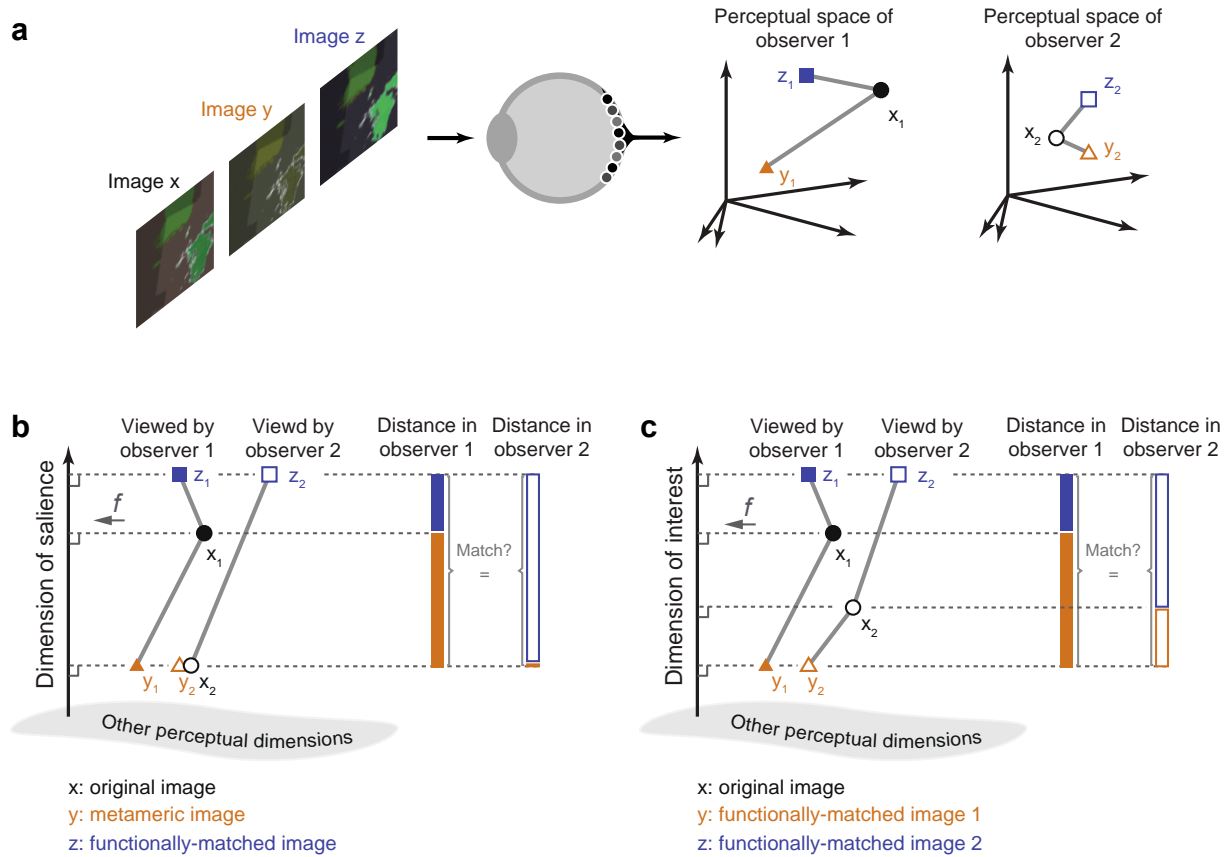

**Figure 2**

A general framework for investigating individual perceptual spaces with functionally matched stimuli.

(a) Each visual image ($x$, $y$, or $z$) is mapped to a perceptual state that is represented as a point in a high-dimensional perceptual space for each observer. The relationships among images form "constellations" in the observer's perceptual spaces, which are generally different from each other.

(b) Perceptual spaces compared based on salience-matched stimuli as references. Image $y$ is a dichromatic metamer for image $x$, whereas $z$ is an image that is different at the retinal level but having salience matched to that of $x$. If the stimuli $y$ and $z$ are both matched in the dimension of salience (denoted by function $f$), we have the following relationships: $(f(z_1) - f(x_1)) + (f(x_1) - f(y_1)) = (f(z_2) - f(x_2)) + (f(x_2) - f(y_2))$ despite that the differential salience values generally differ between the observers, $f(x_1) - f(y_1) \neq \Box(x_2) - f(y_2)$ and $f(z_1) - f(x_1) \neq \Box(z_2) - f(x_2)$. Having these relationships confirmed, we can quantify the difference for the stimulus $x$ between the two observers by $f(x_1) - f(x_2)$ by using images $y$ and $z$ as references for comparing the two different perceptual spaces. The figure shows a simplified situation in which $x_2 = y_2$ which is the case when observer 2 has a complete dichromatic vision. These ideas are supported by the previous simulations and psychophysical experiments (Tajima and Komine, 2015).

(c) The same as b but generalized to arbitrary perceptual dimensions instead of the visual silence. Note that, in general, we do not always have complete metamers, but still have stimuli that are functionally matched between the observers (i.e., $f(y_1) = f(y_2)$ and $f(z_1) = f(z_2)$).

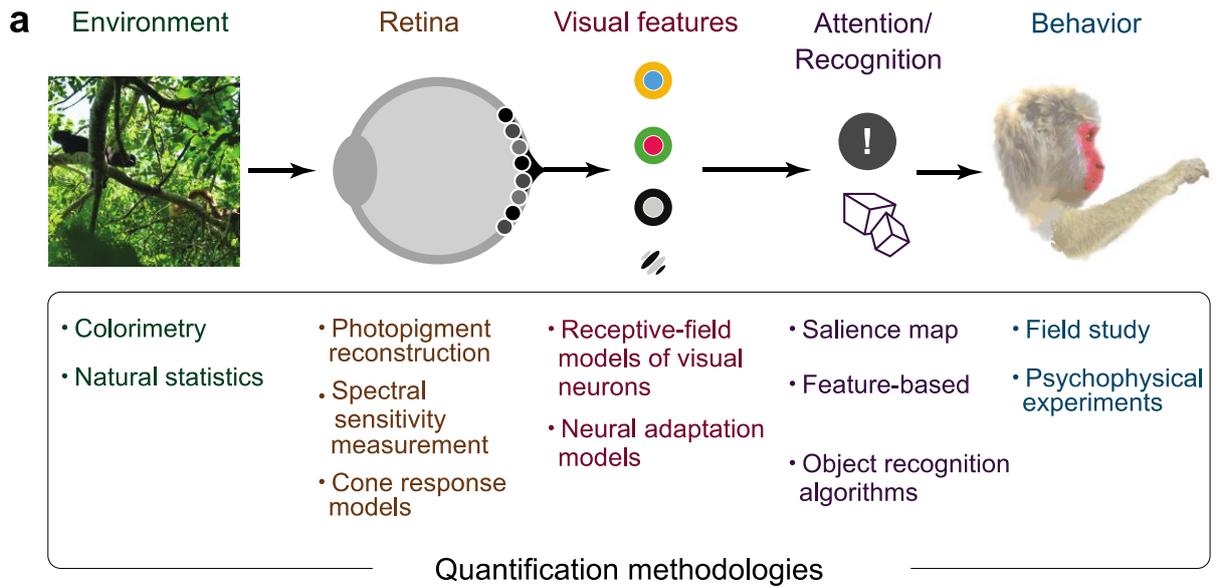

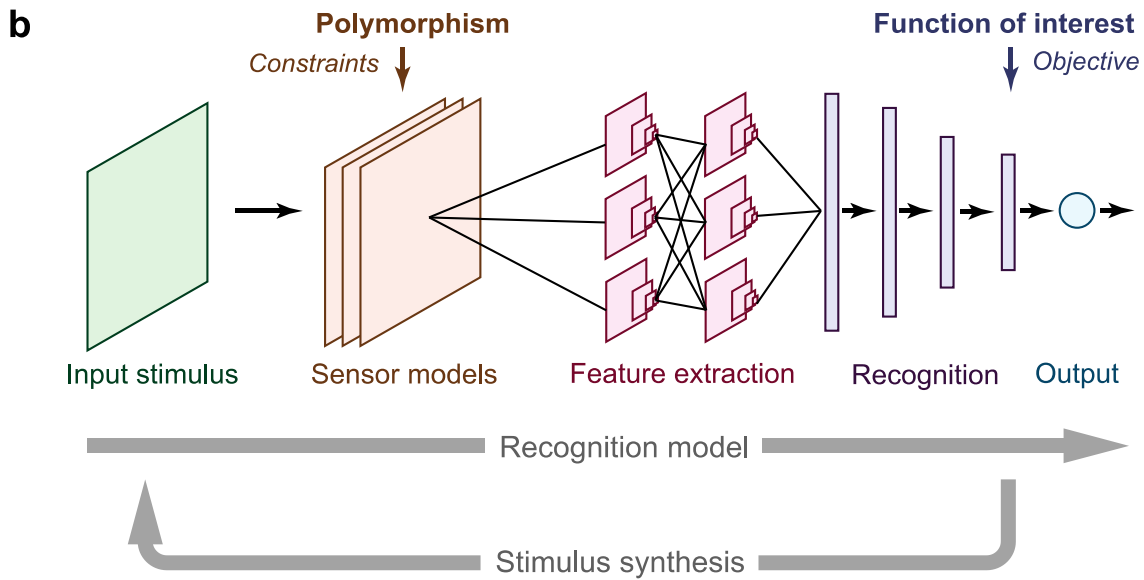

**Figure 3**

Overview of the integrative approach.

(a) (Top) An overall model linking the polymorphism in retinal cone pigments and the behavior. (Bottom) Quantification and modeling methodologies at different stages of the model.

(b) A schematic of the corresponding machine vision and stimulus synthesis processes. The rectangles and circle represent the layers and node within a hierarchical neural model such as a convolutional neural network for object recognition.